# Thinging as a Way of Modeling in Poiesis: Applications in Software Engineering

**Sabah Al-Fedaghi**
**Computer Engineering Department**
**Kuwait University**
**Kuwait**
**sabah.alfedaghi@ku.edu.kw**

*Abstract*—From a software design perspective, a clear definition of design can enhance project success and development productivity. Even though the focus is on software engineering, in this paper, we view the notion of design from the wider point of view of *poiesis*, the field of the study of the phenomena of creation and production of the artifacts. In poiesis, design operates through the medium of modeling. According to several sources, there is as yet no systematic consolidated body of knowledge that a practitioner can refer to when designing a computer-based modeling language. Modeling languages such as UML are practice-based and seldom underpinned with a solid theory—be it mathematical, ontological or concomitant with language use. In this paper, we propose adopting a recent addition to the diagrammatic languages, the thinging machine (abbreviated TM), as a design language in the general area of Poiesis and we exemplify TM by applying it to software engineering design. We show intermediate steps of design that led to producing a TM model for a case study. The case study is taken from a source where a full UML-based design was given. Contrasting the models produced by the two methodologies points to the viability of TM as an integrating and unifying modeling language in the design field.

## I. Introduction

According to Wright [1], "The body of research pertaining to learning to *design* software systems is scarce, with most research efforts studying expert designers and/or comparing novices and experts." Eckerdal et al. [2] found that the majority of graduating computer science students cannot design a complex software system due to several reasons, including the inability to view the system as a whole. According to Ralph and Wand [3], "it is often that information systems academics did not have a well-defined notion of the concept of *design* in the software and IS context; yet, surprisingly, it seems no generally-accepted and precise definition of design as a concept is available." Accordingly, it seems that there is a need for a well-defined and teachable body of knowledge about the principles underlying software systems and the processes used to create them [4]. From a software design perspective, a clear definition of design can enhance software project success and software development productivity [3].

In software engineering, design refers to the iterative analytic transformation of the requirements and specifications for a product into a working system [5]. Software design is the second phase in the development life cycle, after requirements analysis and specifications. "[However], when design is viewed only as a component in the software life-cycle rather than a pervasive activity throughout the development process, aesthetics and usability become far less important than the correctness of the design relative to a given set of specifications." [1]

Ralph and Wand [3] reviewed the notion of design in the literature and identified 20 common concepts: design as process, as creation, as planning, as an activity, and so on. They suggested that the design is a "specification of an object, manifested by an agent, intended to accomplish goals, in a particular environment, using a set of primitive components, satisfying a set of requirements, subject to constraints" [3]. In his PhD thesis, Wright [1] offers a review of design in software engineering and lists a variety of definitions of design from a number of design philosophers (e.g., [6]).

Motivated by the aforementioned difficulties in the notion of design, in this paper, we take a different approach to the notion of design in software engineering, viewing it from a wider perspective (i.e., from the point of view of the design *per sé*) and supplementing this generality with a modeling language, the thinging machine (abbreviated TM), as a specific design tool. Accordingly, software design is part of a field of problem-solving thinking skills that is unique with respect to other abilities such as mathematical or linguistic capabilities. After introducing such an approach, we return to software design and apply TM to a design problem in software engineering.

## II. Design at Large

According to Vial [7], there is a way to express ideas other than by language and other than by mathematical notation in design research: "There exists a designedly way of thinking and communicating that is both different from scientific and scholarly ways of thinking and as powerful as scientific and scholarly methods of enquiry, when applied to its own kinds of problems" [8]. "Thus design activity is not only a distinctive process, comparable with but different from scientific and scholarly processes, but also operates through a medium, called modeling, that is comparable with but different from language and scientific notation [8]. *Design thinking* considers the activity of design as a process of thinking or conceptualization. *Conceptualization* is the conception of a project in terms of the production of an idea that structures the creative process transposed into a scenario. Conceptualization also aims to feed projects with a philosophical reflection on and analysis of uses and social practices. *Design doing* considers the making of a project in progress and involves scenarios taking shape through models. Design is therefore a matter of modeling. While an artist produces artwork, a scientist builds theories and experiences and a philosopher creates concepts, a designer designs and manufactures models. "Designers think in forms …. [and] the act of design is an act of modeling thought" [7].

### A. Poiesis: The Field of Design

Poiesis is the study of the phenomena of creation and production. It includes "design doing of the artefacts and the experience, sensibility and skill that goes into their production and use" [9]. According to Archer [9], the term has been confiscated by the arts and humanities; it could also have been called "technology" (here, technology in the sense of giving rise to material artifacts).

Poiesis, as interpreted by Heidegger [10], refers to productive behavior in whose production the designer projects a model of the thing. In this Heideggerian notion of design/production, the designer not only diagrams the material or software, such that it embodies the projected model, but in so doing liberates the material or software from its dependence on the designer until, as it eventually materializes, it obtains an independent being-in-itself [10].

Poiesis is the field of what is produced by design. It asks, "In what way is design a practice of creation? What is the creative work of a designer? In what sense can one speak of an 'act of design' and how does it work?" [7]. Poiesis involves two intimately intertwined and inseparable dimensions: the dimension of the act of thinking and the dimension of the act of making [7].

### B. Design Problems: Typically Ill-Defined

It is now widely recognized that design problems are ill-defined, ill-structured, or "wicked" [11]. They are not the same as the puzzles that scientists, mathematicians and other scholars craft for themselves. They are not problems for which all the necessary information is available. They are therefore not exhaustively analyzed, and there can never be a guarantee that a "correct" solution-focused strategy is clearly preferable to continue analyzing a "problem", but the designer's task is to produce "the solution" [12].

An ill-defined problem is one in which the requirements as given do not contain sufficient information to enable the designer to arrive at a means of meeting those requirements by transforming, reducing or superimposing the given reformation. [8]

A central feature of design activity, then, is its reliance on fairly quickly generating a satisfactory solution, rather than on any prolonged analysis of the problem. To use Simon's [13] term, it is a process of "satisficing" rather than optimizing—producing any one of what might well be a large range of satisfactory solutions rather than attempting to generate the one hypothetically optimum solution. [12].

What designers tend to do, therefore, is to seek or impose a "primary generator", which both defines the limits of the problem and suggests the nature of its possible solution [12].

### C. Design in Software Engineering

According to Vial [14], design is a process. Rephrasing Vial's [14] proclamation, design is not a phase in the software development life cycle; rather, it is software engineering that is an area for applying design. Design is the idea of "enchanting" software. A software system's design is not a being but an event; not a thing but an impact; not a property but a repercussion; not conceiving "things that are" but conceiving "things that happen" [14].

Designing a software system means first creating forms (e.g., diagrams). When no harmony of forms exists, no design can take place. Software engineering is an operational discipline (concerned with doing or making) that deals with software systems (the programming artifacts) themselves and the sensibility, invention, validation and implementation that arise in their production and use [8].

Design awareness in software engineering means the ability to understand and handle ideas that are expressed through the medium of "softwar-ing." Modeling is the language in which such ideas may be expressed.

### D. Modeling

The use of models to understand and shape the world is a foundational technique that has been in use since ancient Greece and Egypt [15]. Modeling is an essential skill to express ideas by means other than language and notation [7]. "Human beings have an innate capacity for cognitive modeling, and its expression through sketching, drawing, construction, acting out and so on, that is fundamental to thought and reasoning as is the human capacity for language" [8].

Although modeling has been employed for ages, "it is fairly new that the form of models is made explicit in modeling languages" [15]. "The essential language of design is modelling" [9]. The model is not only a tool, a method or a stage; it is a place where one projects ideas for the future. It is anticipation through diagramming [9]. For modelers, the world is "a project and not just an object that must be described, whose causes must be explained or whose meaning must be understood" [16].

The concept of modeling takes on a philosophical and anthropological aspect "because it is not only a technique of representation, as it can be in other design disciplines such as [classical] engineering. It is a place for the development of an ideal that is taking form. Through [the model], the modeler creates ideas, but these ideas are not the 'concepts' of science or philosophy, nor the 'affects' of art" ([7] referencing [17]). This form of ideal is an executable must-be and has the form of both an ideal and an operational concept [7].

A model is a representation of something that captures, analyzes, explores and transmits those ideas. "The modeling of ideas in the Design area can be conveyed through a variety of media such as drawings, diagrams, physical representations, gestures, algorithms—not to mention natural language and scientific notation" [8].

## III. Modeling Languages

According to Bohnke et al. [18], design languages initially emerged from the fields of biology (for the modeling of plant growth [19]) and architecture (for space and pattern generation [20]), and design languages have been applied to the automated synthesis of a variety of engineering, as well as industrial product designs. Among them are coffeemakers [21], houses [22], gears [23], chairs [24] and satellites [25]. Design languages may be differentiated into the three distinctive kinds of design language representations: string-based, shape-based and graph-based [18]. Note that the focus in this paper is on diagrammatic languages. According to Tversky [26], "Design without drawing seems inconceivable. . . Sketches can reveal thought, then. . . [and] they can reveal the elements or segments of construction or of thought in a particular domain."

For many popular modeling languages, such as UML, AADL, or Matlab/Simulink, research and industry have produced useful analyses and transformations. These rely on making the constituents and concerns of languages machine-processable [15]. Modeling languages is still much more of an art than a science [27]. "There is as yet no systematic consolidated body of knowledge that a practitioner can refer to when designing a computer-based modeling language" [27]. Contemporary software engineering modeling tends to rely on general-purpose languages such as UML. "However, such languages are practice-based and seldom underpinned with a solid theory—be it mathematical, ontological or concomitant with language use" [28].

Hölldobler et al. [15] presented an overview of the current state of the art on modeling languages, including programming languages and design specific languages. They concluded that

Programming languages in general, SQL, XML, and the Unified Modeling Language (UML) in particular have been created to enable highly precise communication. Despite these efforts, it is clear that researchers and practitioners of many domains are dissatisfied by solving domain-specific problems with general purpose languages or unified languages that try to cover everything. . . these languages suffer from not being very domain-oriented.

Bohnke et al. [18] focused on design languages and expected that the higher level of abstraction offered by graph-based design would ease the tasks of model updating in early design phases.

In this paper, we propose adopting a recent addition to diagrammatic modeling languages, TM, a design language in the general area of poiesis (beyond software engineering). One reason for such a proposal is the observation that TM can have a wide variety of applications, including phone communication [29], physical security [30], vehicle tracking [31], intelligent monitoring [32], asset management [33], information leakage [34], engineering plants [35], inventory management processes [36], procurement processes [37], public key infrastructure network architecture [38], bank check processing [39], wastewater treatment [39], computer attacks [40], provenance [41], services in banking industry architecture networks [42] and digital circuits [43]. TM has also been used to model poetry [44], storytelling [45], philosophical thoughts [46] and privacy [47].

We suspect that TM is somehow related to a deep structure of modeling. In particular, one characteristic of TM is its use of only five generic verbs and the claim that this is sufficient to describe all processes.

We claim that TM is a design language that allows the construction of a variety of designs; nevertheless, we focus on software design and show steps of design leading to TM model.

## IV. Thinging Machines

The TM model starts with things/machines, or *thimacs*, which populate a world that is itself a thimac (we call it a system). Every part of this world is a thimac. forming a chain of thimac-ing. A unit of such a universe has dual being as a *thing* and as a *machine*. A thing is what is created, processed, released, transferred and/or received. A machine is what creates, processes, releases, transfers and/or receives. We will alternate between the terms thimac, thing or machine, according to the context.

The term "thimac" designates what simultaneously divides and brings together a thing and a machine. Every thimac either appears in the system by creation or by importation from outside the system. They are the concomitants (required components) of the system and form the current fixity of being of any system that continuously changes from one form (thing/machine) to another.

Accordingly, the existence of a thimac depends on its position in the larger system; either a thing that flows in other machines or a machine that handles a flow of things (create, process, release, transfer and receive). It brings together and embraces both "thingishness" and "machineness." A thimac may act in the two roles simultaneously.

A thing (ignoring its mechanical nature) flows in an abstract five-dimensional structure that forms an abstract machine called a TM, as shown in Fig. 1, where the elementary processes are called the *stages* of a TM. A TM can be put into the form of the input–process–output model (Fig. 2).

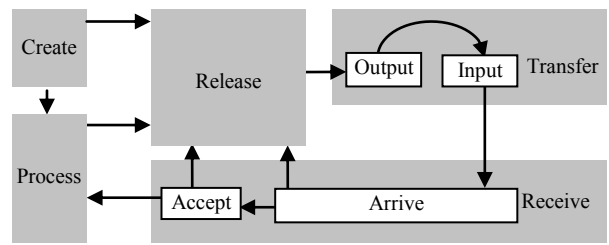

**Figure 1. Thinging machine.**

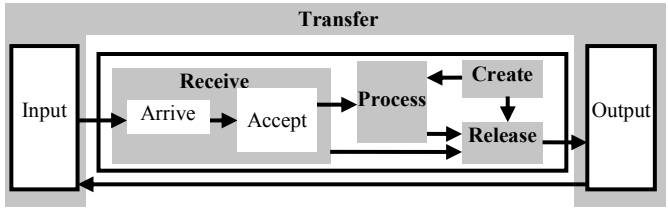

**Figure 2. Another description of a TM.**

In the TM model, we claim that five processes of things exist: things can be created, processed, released, transferred, and received. These five processes form the foundation for modeling thimacs.

Among the five stages, flow (solid arrow in Fig. 1) signifies conceptual movement from one machine to another or among the stages of a machine. The TM stages can be described as follows:

**Arrival**: A thing reaches a new machine.

**Acceptance**: A thing is permitted to enter the machine. If arriving things are always accepted, arrival and acceptance can be combined as the **receipt** stage. For the purpose of simplification, the examples in this paper assume a received stage.

**Processing** (change): A thing undergoes some kind of transformation that changes it without creating a new thing.

**Release**: A thing is marked as ready to be transferred outside of the machine.

**Transference**: A thing is transported somewhere from/to outside of the machine.

**Creation**: A new thing is born (created) in a machine in an analogy to the dynamic creation of objects in the UML. The term *create* comes from creativity with respect to a system (i.e., constructed things from already created things, or emergent things appearing from somewhere).

Additionally, the TM model includes **memory** and **triggering** (represented as dashed arrows) relations among the processes' stages (machines).

A machine creates in the sense that it "to-finds/originates" a thing in a sense of bringing it into the system and that the machine becomes aware of it. Creation can be used to designate "bringing into existence" in the system because what exists is what is found. For simplification's sake, we may omit creation when drawing a machine.

We can visualize the machine as a soap bubble (thus, we have a complex of bubbles) that includes the thing and its flow environment. A thing can only flow from one stage to another in a machine. The thing itself is a bubble for other things. The basis bubble includes only the creation stage.

**Example**: Consider the UML activity diagram of a simple ATM shown in Fig. 3. Fig. 4 shows the corresponding TM model, which shows different thimacs. Fig. 4 shows a static model of an ATM. First, the card is inserted (circle 1 in the figure) to flow to the ATM (2) to be processed (3) to extract the card number (4). Also, the PIN is input (5) and flows to the ATM (6) where it is compared with the card number (7).

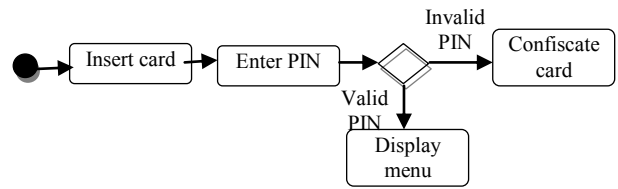

**Figure 3. An ATM example** *(Redrawn from https://creately.com/blog/diagrams/uml-diagram-types-examples/).*

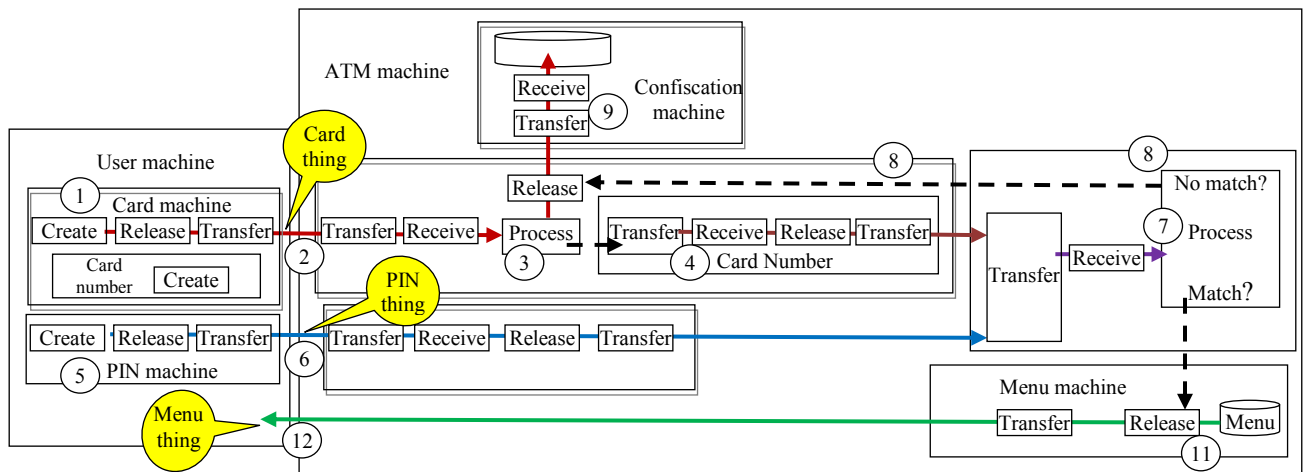

**Figure 4. TM model of the simple ATM example.**

If the card number and the PIN do not match, the triggering mechanism (8, dashed arrow) releases the card to be confiscated (9). If they match, a trigger releases a menu (10) and sends to the user (11).

To develop a dynamic model, events are superimposed on the static model. Events in TM are machines that encompass at least three submachines: time, region and the event itself. For example, Fig. 5 shows the event *the card is confiscated*.

Accordingly, we select the following events.

Event 1 ($E_1$): The card is inserted and received by the simple the machine.

Event 2 ($E_2$): The PIN is entered into the machine.

Event 3 ($E_3$): The card number is compared with the PIN.

Event 4 ($E_4$): The card is confiscated.

Event 5 ($E_5$): The menu is sent to the user.

Figure 6 shows the behavior of the simple ATM in terms of its selected events.

Note that the TM model is built in a search for machines on one side and identifies things and their flows on the other side. The TM model is a network of thimacs in which the designer perceives the two forms of being—things and machines. A designer's thimac-based thinking excludes other modes of thinking such as procedural and object-oriented modes of thinking. "Every person is conscious of a train of thought being immediately awakened in his imagination analogous to character or expression of the original object" [48]. Each "thimac thought" follows another, connected by flows of things whereby thimacs change their being from the subject form to an object form and vice versa. Thimac-based modeling is a thinking style that involves how one organizes thoughts; it is a "conscious system of design." [49].

The thimac thinking style can be applied in any domain. Consider mathematics: a set can be conceptualized as a thimac that handles things called elements that flow through and into and out of the thimac system. Handling here refers to the acts of transferring, receiving, processing, creating, and releasing elements of the set. A function such as f: X → Y can be represented in terms of a map between thimacs, conceptualized not in terms of space and movement, but in terms of flows in which the values of input things are changes in the output things. Such a view may have some application in teaching mathematics [50].

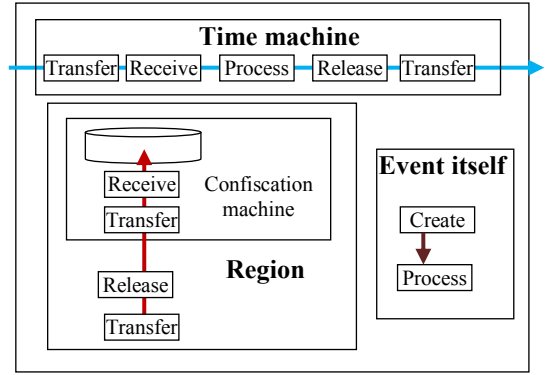

**Figure 5. The event "the card is confiscated."**

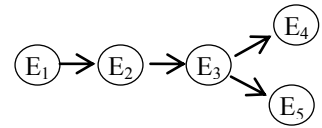

**Figure 7. Chronology of events.**

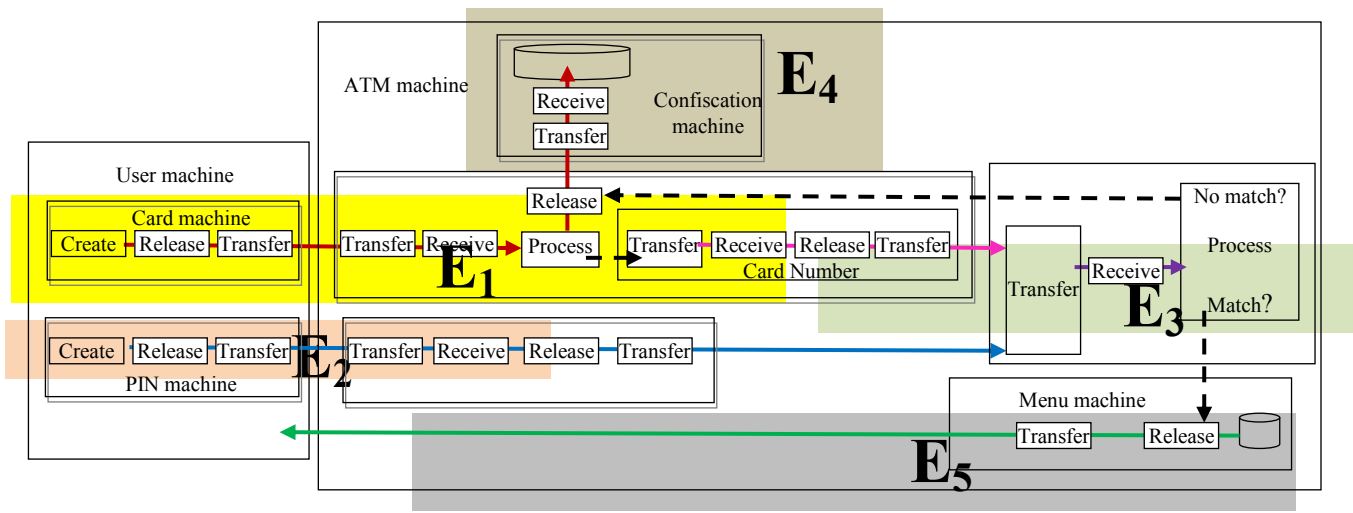

**Figure 6. Events in the TM model of the ATM example.**

## V. CASE STUDY

Gomaa [51], in his book *Software Modeling and Design*, presents several design case studies. One of these case studies is the design of client/server software architecture: a banking system. The first part of the case study is described by Gomaa [51] as follows.

A bank has several automated teller machines (ATMs) that are geographically distributed and connected via a wide area network to a central server. Each ATM machine has a card reader, a cash dispenser, a keyboard/display, and a receipt printer. By using the ATM machine, a customer can withdraw cash from either a checking or savings account, query the balance of an account, or transfer funds from one account to another...

### A. *Initial Steps*

Gomaa [51] used 37 diagrams (mostly UML) in developing the design of the problem. To approach the design from the TM side, the given English description is examined and some words are highlighted, with some parts separated from each other, as shown in the following highlighted portion.

A transaction is initiated when a customer inserts an ATM card into the card reader. Encoded on the magnetic strip on the back of the ATM card are the card number, **the start date**, and the **expiration date**. The system **validates** the ATM card to determine:
- the expiration date has not passed,
- **the user-entered personal identification number, or PIN, matches the PIN maintained by the system,**
- **the card is not lost or stolen.**

Thus, applying initial thimac thinking, we try to search for machines. A rough initial TM is produced, as shown in Fig. 7. Fig. 8 shows attempts to identify things and their flows. The backbone of the system is recognized as a flow of cards that triggers (i) the extraction of card number and expiration data, (ii) the flow of card number to be compared with the PIN; and (iii) the card number flow to the bank to check for lost or stolen cards.

### B. *First version*

Hence, the first version of the design is produced as shown in Fig. 9, which can be described as follows:

- An ATM card is inserted into the card reader (circle 1), received and processed (2).
- The card number is extracted from the card (3).
- A PIN is input (4) that is received and compared with the card number (5).
- The card is rejected and ejected (6).
- The card number is sent to the bank (7), where it is processed. The response flows to the ATM (8).
- The response is negative, so the card is confiscated (9).
- The expiration date is extracted (10) from the card, and processed against the current date (11).
- The card is rejected and ejected (12).

For simplification, we omit modeling the display on the ATM screen.

This is followed by examining Gomaa's [50] original English description again. It is discovered that the part "The customer is allowed three attempts to enter the correct PIN; the card is confiscated if the third attempt fails" was not included in the first version of the TM model. Accordingly, Fig. 9 is amended to incorporate this missing part of the model, producing Fig. 10.

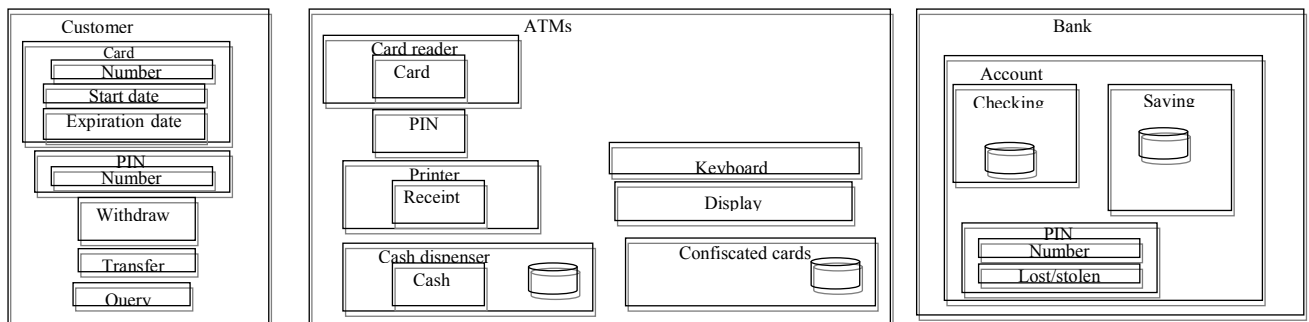

**Figure 7. First draft that identifies machines.**

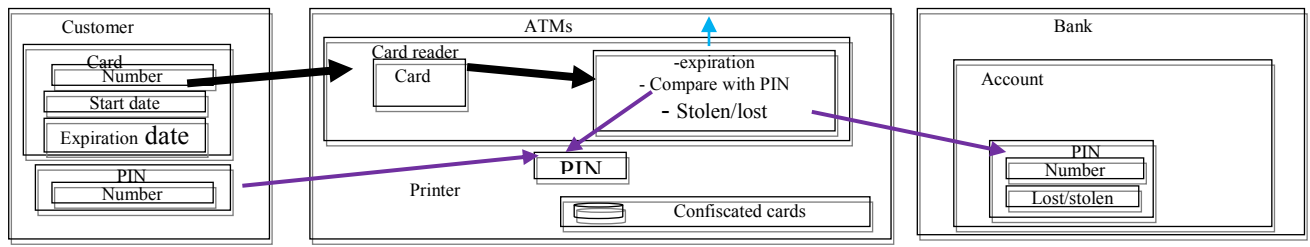

**Figure 8. Identification of main flows.**

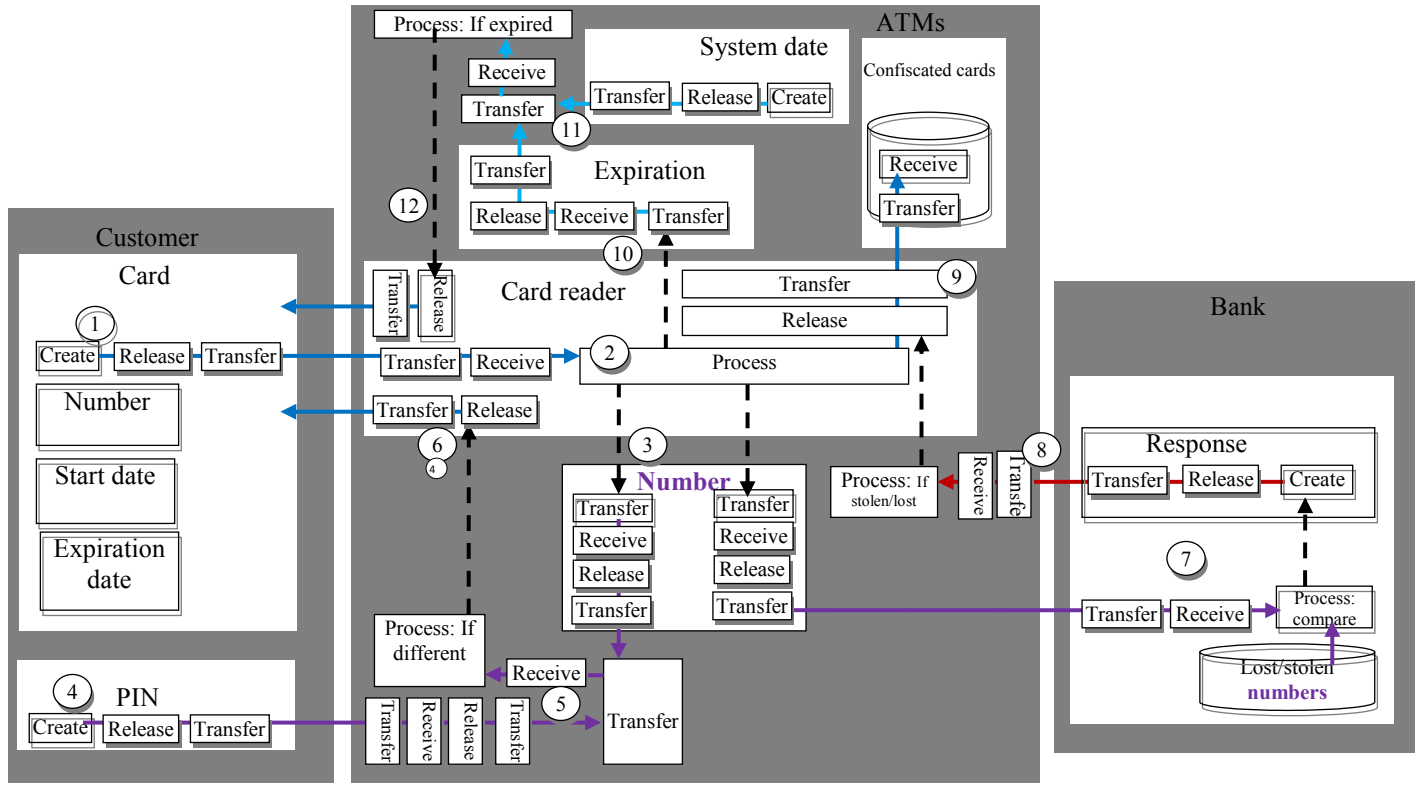

**Figure 9. First version of the TM model.**

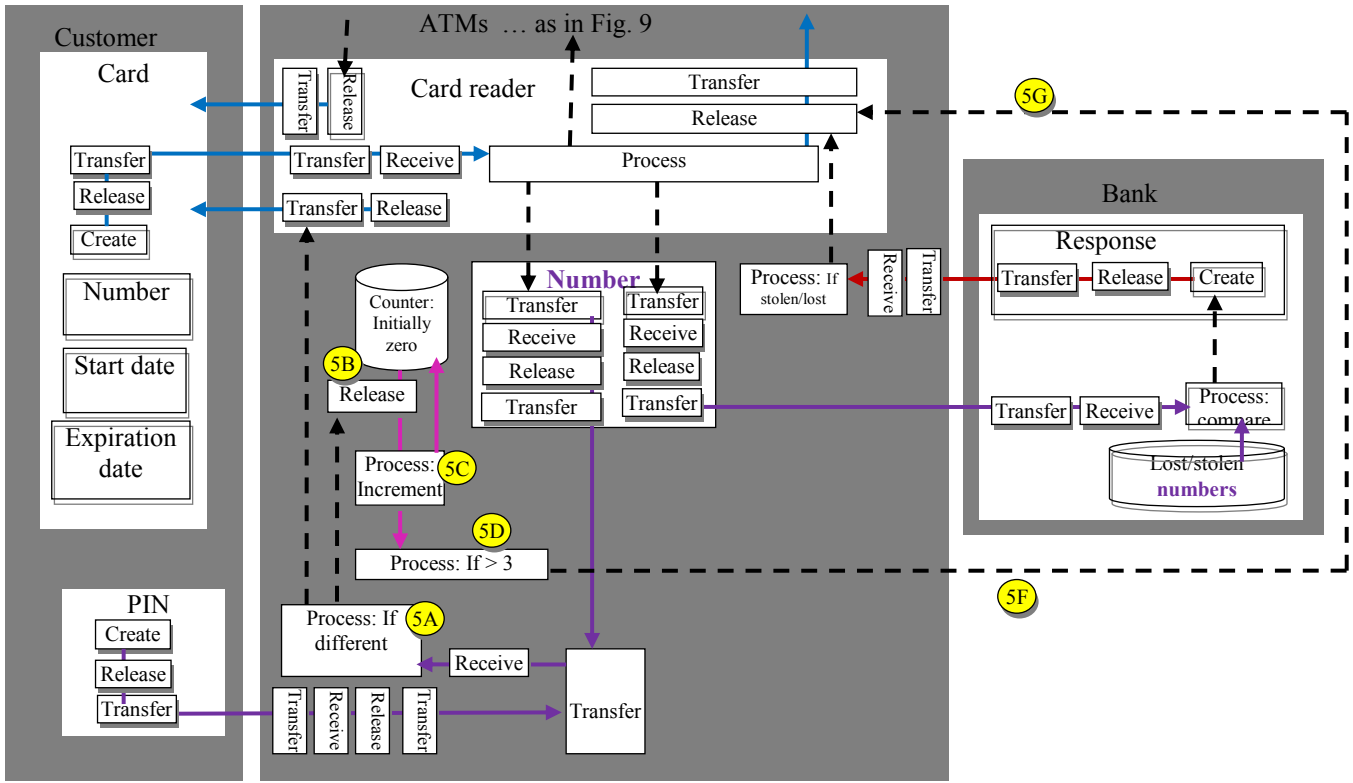

**Figure10. Fixing the omitted part in the second round of checking, the design is the TM model of the ATM example.**

In Fig. 10, the addition to Fig. 9 is inserted at circle 5A, when the card number and the PIN are compared, and a discrepancy is found. There, we add a counter (5B) that is incremented and updated (5C) each time the comparison produces a negative result. If this failure occurs more than three times (5D), the card is confiscated (5E and 5G).

### C. Behavior

Continuing with Gomaa's [51] case study, we model the behavior of the system by identifying events, as discussed in section IV. The TM model is the basis in which activities are synchronized. Transferring, releasing, receiving, processing, and creating things are changes in the flow of things and a hierarchy of events can be built out of them to visualize "larger" events.

At the semantic level, we identify these larger events to divide the system into chunks of knowledge regions that are more meaningful to the human mind. Such a selection is a design issue.

In our study case, we identify the following events (See Fig. 11).

Event 1 ($E_1$): The card is inserted and received by the ATM machine.

Event 2 ($E_2$): The card is processed.

Event 3 ($E_3$): The card number is extracted.

Event 4 ($E_4$): The PIN is input.

Event 5 ($E_5$): The PIN and the card number are compared, resulting in ejection or confiscation of the card.

Event 6 ($E_6$): The card number is sent to the bank where it is checked for being stolen or lost.

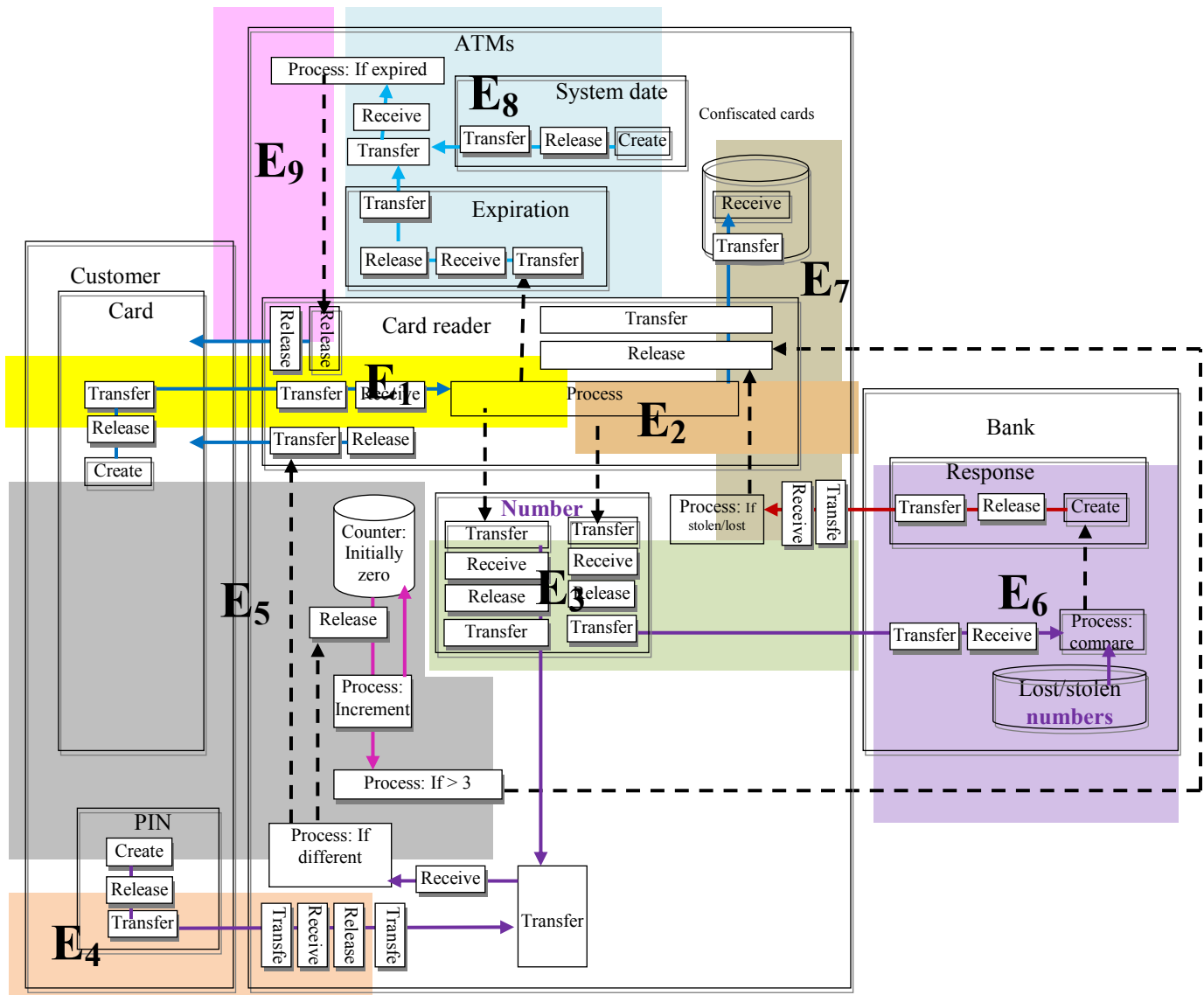

**Figure 11. Events in the TM model of the ATM example.**

Event 7 ($E_7$): The card is confiscated.

Event 8 ($E_8$): The expiration date is extracted and compared with the current date.

Event 9 ($E_9$): The card is ejected.

Fig. 12 shows the behavior of the system in terms of the chronology of these events.

### D. *The Rest of the Solution*

Accordingly, it is clear where we can link the first part of the ATM case study with the second part. There are three points of contact, as shown in Fig. 13.

- When the PIN is equal to the card number
- When the card is not lost or stolen, and
- When the expiration date has not passed.

When these three conditions are satisfied, then we are in the second part of the design process which, as in the first part, starts with an English description,

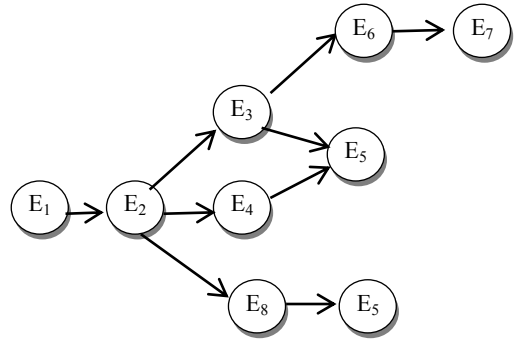

**Figure 12. Chronology of events in the first part of the ATM machine model.**

The whole design process used in Part 1 of the case study is repeated. The process is a straightforward repetition of the design in the first part, and for space considerations, it will not be shown in this paper.

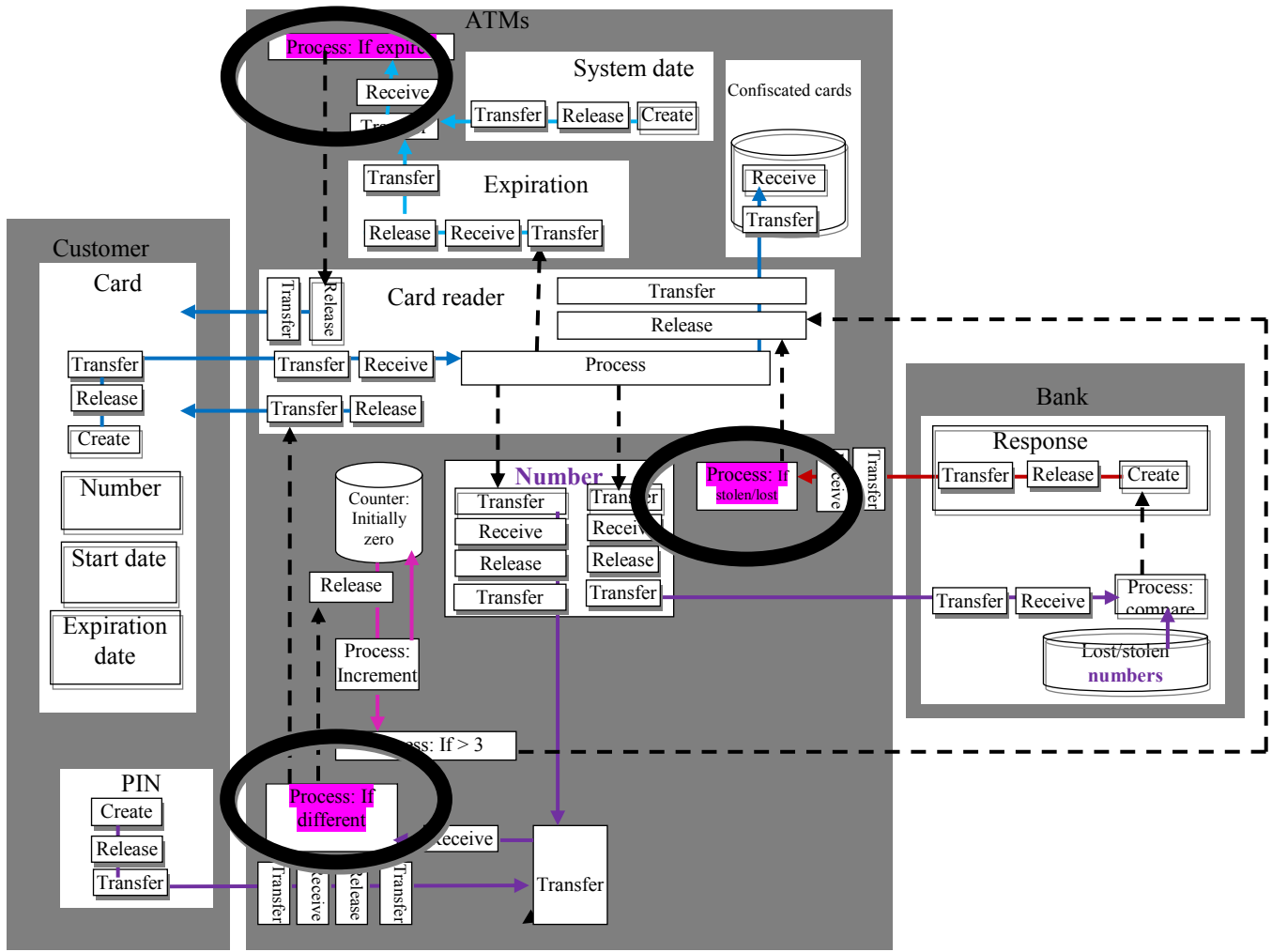

**Figure 13. Identifying the "joints" between the first and second parts.**

### E. What are Instances?

At this point, the question may be raised about the notion of *instance* in TM, as with the distinction between object and class in UML. In ontology studies, this distinction is called universals and particulars and is characterized by taking the relation of instantiation [52]. "For example, Pavarotti, the Italian tenor, is an instance of 'person', but he cannot himself be instantiated. (This characterization of the concept of universal is admittedly imprecise since it does not, for instance, clarify whether sets, predicates and abstract entities should be considered as universals or not.)" [52].

In TM, instantiation refers to a time machine that "engulfs" the timeless static TM machine, converting it to a three-dimensional model by adding time.

An instance is an event that refers to facts (not history). Take for example the thimac card in Fig. 10. The *card machine* in the figure seems to imitate a class in UML with its attribute number, start data and expiration data. A card instance (object) in UML includes a tuple that gives values to the attributes.

This way of constructing an instance of the card can be modeled in TM as follows. First we have to align the TM card machine with the above description by specifying how the card is created, as shown in Fig. 14. We ignore the expiration date because its treatment is similar to the start date. Fig. 14 gives finer details of creating a card by implanting values in it though an integer machine. The integer machine produces four types of things:

- A positive integer string (say, 9 digits) that is exported to the card number.
- A 2-digit integer string between 1 and 31 that is exported to day.
- A 2-digit integer string between 1 and 12 that is exported to month.
- A 4-digit integer string between, say, 2000 and 2050 that is exported to year.

An instance of card (data) is created by the large event that includes the events shown in Fig. 15. Of course, an event involves time, which is not shown in Fig. 15. Every instance has its timestamp. The purpose here is to illustrate how instances are types of events in TM.

Note that creating instances is an integrated part of the behavior of the system superimposed on the static TM model. So, instances represent waves of events over the diagram. For instance, in the current example, a card being (conceptually) created means the appearance of a tuple, say, (123456789, 01-06-2015, 01-06-2019) that goes through different events of the behavior of the system, such as extracting 123456789 until finishing the transaction. Then another wave, related to a different card (different instance), starts occurring through the same path of events. This eliminates the need to model another diagram for instances, as in the case of class diagram and object diagram in UML.

For example, consider the class and object diagrams given in [53] and shown in Fig. 16. Fig. 17 shows the corresponding TM model. Instances (objects) are regarded as repeated events over the same region.

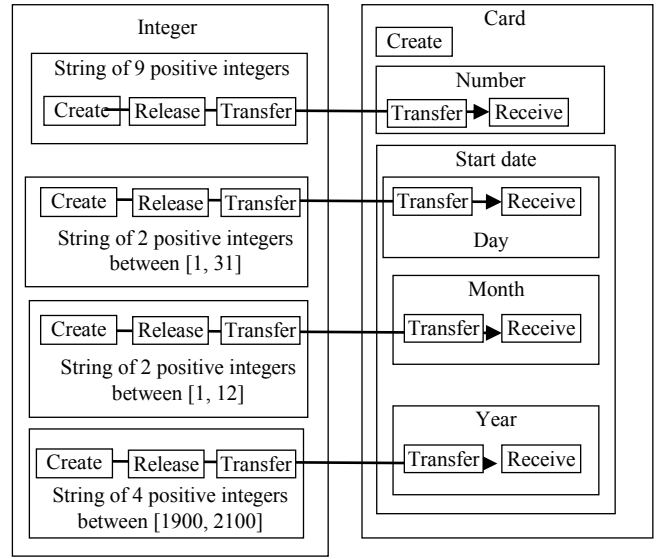

**Figure 14. Card machine.**

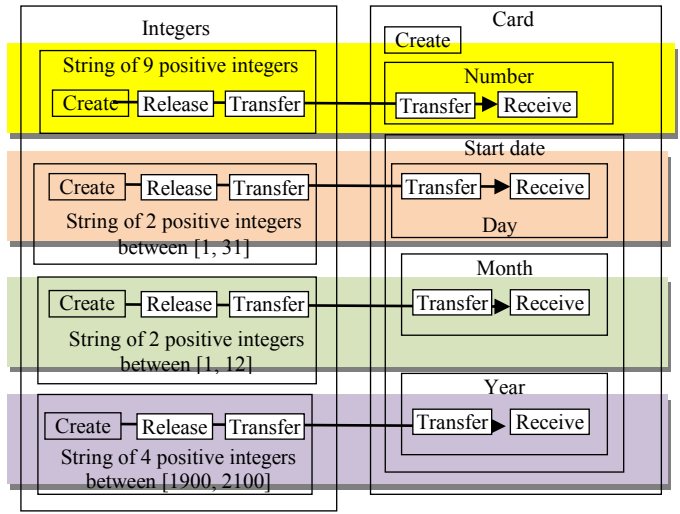

**Figure 15. Creating an instance in the card machine.**

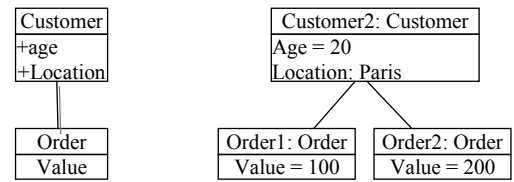

**Figure 16. Sample of class diagram (left) and object diagram (right) (Partially redrawn from [53]).**

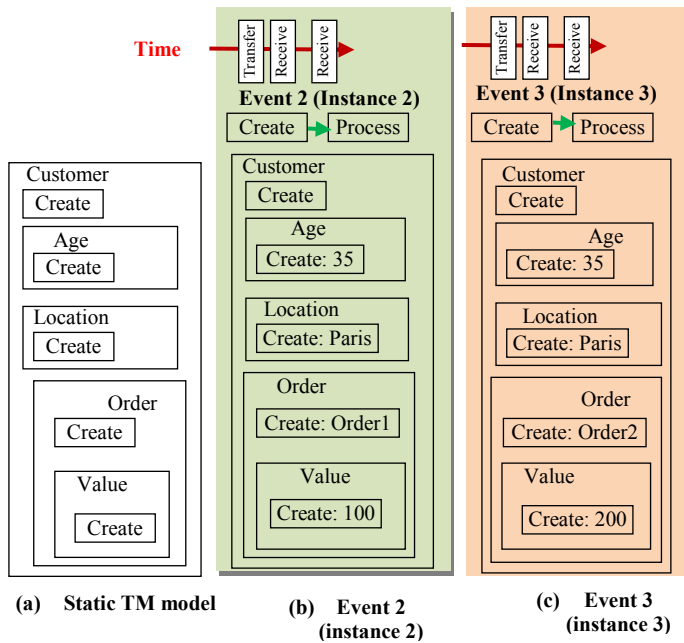

**Figure 17. Superimposing events over the TM model as instances.**

## VI. CONCLUSION

According to Penrose [54], "There seem to be many different ways in which different people think—and even in which different mathematicians think about their mathematics." Likewise, in design, a variety of methods of thinking apply to design. In this paper, we proposed adopting TM as a design language in the general area of poiesis.

We speculate that TM thinking is a thinking style and claim that thinking and diagramming this way leads to a viable design methodology. In the context of TM, we assert that TM can provide a basis for thimactic thought and that elements and segmented construction of thought appear as elements flow through machines. The first phase of design includes identifying machines, then things and their flows.

We suspect that TM is somehow related to a deep structure of modeling, especially with regard to its use of only five generic verbs and the claim that these are sufficient to describe all processes. These five processes define the thimac: the thing and the machine. In a way, thimactics is a diversion from the dualism of separation of subject and object. The machine part of the thimac is the subject and its thing part is the object. More research is needed to clarify this area.

TM is a design language that allows the construction of a variety of designs. Nevertheless, from the practical side, we have focused on software design and showed intermediate steps of design that led to producing a TM model for a case study. We introduced a case study taken from a source in which 37 diagrams were used to design a system. In actual advanced software engineering classes, students completely lose a reasonable grip on the solution when they get lost in this jungle of disintegrated diagrams. Teaching the same problem using TM theory seems to ease this problem.